# The Robinson Gravitational Wave Background Telescope (BICEP): a bolometric large angular scale CMB polarimeter

K. W. Yoon[†a], P. A. R. Ade[b], D. Barkats[a], J. O. Battle[c], E. M. Bierman[d], J. J. Bock[a,c], J. A. Brevik[a], H. C. Chiang[a], A. Crites[a], C. D. Dowell[c], L. Duband[e], G. S. Griffin[a], E. F. Hivon[f], W. L. Holzapfel[h], V. V. Hristov[a], B. G. Keating[d], J. M. Kovac[a], C. L. Kuo[c], A. E. Lange[a], E. M. Leitch[c], P. V. Mason[a], H. T. Nguyen[c], N. Ponthieu[g], Y. D. Takahashi[h], T. Renbarger[d], L. C. Weintraub[a], D. Woolsey[h]

[a] California Institute of Technology, Pasadena, CA 91125, USA
[b] Dept. of Physics and Astronomy, University of Wales, Cardiff, CF24 3YB, Wales, UK
[c] Jet Propulsion Laboratory, Pasadena, CA 91109, USA
[d] Dept. of Physics, University of California at San Diego, La Jolla, CA 92093, USA
[e] SBT, Commissariat à l'Energie Atomique, Grenoble, France
[f] Infrared Processing and Analysis Center, Pasadena, CA 91125, USA
[g] Institut d'Astrophysique Spatiale, Université Paris-Sud, Orsay, France
[h] Dept. of Physics, University of California at Berkeley, Berkeley, CA 94720, USA

## ABSTRACT

The Robinson Telescope (BICEP) is a ground-based millimeter-wave bolometric array designed to study the polarization of the cosmic microwave background radiation (CMB) and galactic foreground emission. Such measurements probe the energy scale of the inflationary epoch, tighten constraints on cosmological parameters, and verify our current understanding of CMB physics. Robinson consists of a 250-mm aperture refractive telescope that provides an instantaneous field-of-view of 17° with angular resolution of 55′ and 37′ at 100 GHz and 150 GHz, respectively. Forty-nine pair of polarization-sensitive bolometers are cooled to 250 mK using a $^4$He/$^3$He/$^3$He sorption fridge system, and coupled to incoming radiation via corrugated feed horns. The all-refractive optics is cooled to 4 K to minimize polarization systematics and instrument loading. The fully steerable 3-axis mount is capable of continuous boresight rotation or azimuth scanning at speeds up to 5 deg/s. Robinson has begun its first season of observation at the South Pole. Given the measured performance of the instrument along with the excellent observing environment, Robinson will measure the E-mode polarization with high sensitivity, and probe for the B-modes to unprecedented depths. In this paper we discuss aspects of the instrument design and their scientific motivations, scanning and operational strategies, and the results of initial testing and observations.

## 1. INTRODUCTION

Numerous experimental efforts in recent years have confirmed that the temperature and polarization anisotropies in the cosmic microwave background (CMB) contain a wealth of information about the evolution and content of the universe. The detailed results of these observations have provided a growing confidence in the inflationary paradigm of the origin of the cosmos. One of the basic predictions of the theory of recombination[1]—the decoupling of the CMB from matter during the formation of atoms at redshift ~ 1100—is that the CMB will be polarized at the level of a few percent of the temperature anisotropy, and that this polarization will have a curl-free, or E-mode, pattern that is correlated with the temperature anisotropy. This polarization signature has now been detected by several experiments over a wide range of angular scales (arcminutes to tens of degrees), providing a heartening and important confirmation of our basic understanding of CMB physics. DASI was the first to report a detection in 2002,[2] followed by WMAP's 1-yr measurement of TE-correlations,[3] and more recently complemented by results from BOOMERANG,[4,5] CBI,[6] CAPMAP,[7] and multi-year analyses from both DASI[8] and WMAP[9], among others.

---

[†] Corresponding author: Ki Won Yoon, 1200 E California Blvd, MC 59-33, Pasadena, CA 91125. Email: kiwon@caltech.edu

| Band Center (GHz) | Band Width (GHz) | No. of Detectors | Beam FWHM (arcmin) | NET$_{CMB}$ ($\mu$K·s$^{1/2}$) |
|---|---|---|---|---|
| 97.7 | 24.4 | 50 | 55 | 480 |
| 151.8 | 42.9 | 48 | 37 | 420 |

Table 1. Measured performance of the Robinson instrument. The NET is for a single polarization sensitive bolometer (PSB).

By verifying the theoretical model and further tightening the constraints on the cosmological parameters, the polarization measurements thus far have laid a strong foundation for tackling an even more elusive goal: measuring the B-mode polarization of the CMB. As much as the temperature and E-mode polarization are a relic of scalar or density fluctuations in the early universe, the B-modes are a direct probe of primordial tensor or gravitational wave perturbations[10]—a generic prediction of inflationary models which describe a rapid exponential expansion of the universe during the first $10^{-38}$ seconds following the initial singularity.[11] So far, inflation has passed every observational test. Yet the critical test still remains: the detection of gravitational wave background (GWB) and its faint imprint on the CMB polarization.

The Robinson GWB Telescope, the first receiver on the BICEP experiment,[12] is a mm-wave bolometric array designed specifically to target the B-mode polarization of the CMB at degree-scale angular resolution, where the signature of the primordial GWB is expected to peak. Such a measurement would be powerful new evidence in support of inflation, and a remarkable probe of the energy scales of the earliest processes in the birth of the universe.

## 2. INSTRUMENT DESCRIPTION

A realistic attempt at measuring the B-modes from the ground requires not only high system sensitivity, but an instrument and observing strategy specifically geared towards exquisite control of systematics. In tailoring the Robinson Telescope for sensitive detection of degree-scale polarization, we have taken care to minimize or avoid possible sources of systematic contaminations. Robinson utilizes field-proven detectors, instrumentation, and mm-wave technologies, albeit combined with a novel optical and operational design, resulting in a reliable, well-characterized, and compact instrument optimized for its inflationary science goals.

### 2.1. Detectors

Robinson uses a pair of silicon nitride micromesh polarization sensitive bolometers[13] (PSBs) at each pixel location, orthogonal in polarization orientations and co-mounted behind shared feeds and cold band-defining filters. A total of 49 such pairs, 25 at 100 GHz and 24 at 150 GHz, are cooled to 250 mK to achieve close to background-limited sensitivities. After adjusting for the relative responsivities, the pairs are summed or differenced to measure the Stokes *I* or *Q* parameters. Because the orthogonal PSBs observe the CMB through the same optical path and atmospheric column with near-identical spectral pass bands, systematic contributions to the polarization measurement are greatly minimized.

Due to its wide intrinsic bandwidth, the same micromesh bolometer design can be used for both 100 GHz and 150 GHz band, in detector housings tuned for band-specific $\lambda$/4 backshort integrating cavities and appropriate mating to the corrugated PSB feed horns. Figure 1 shows a typical Robinson PSB prior to being mounted. The optically active absorber mesh area is 4.5 mm in diameter, metalized in the direction indicated and suspended along the perimeter by low thermal conductance support legs. A neutron transmutation-doped germanium (NTD Ge) thermistor is mounted at the edge of the absorber, placed outside the coupling area to the incoming guided wave to minimize crosspolar response.

Properly tuned, NTD Ge bolometers can offer background photon noise-limited sensitivity in low loading conditions.[14] The thermal conductance $G(T)$ between the thermistor and the 250 mK bath is tailored for the expected total loading to limit phonon noise. The conductance of the Robinson PSBs is dominated by the signal leads and a pair of adjacent metalized legs, which can be laser-trimmed to change the final conductance to {40, 50, 67}% of its full value. Prior to deployment, the PSBs for both bands were adjusted to a target value of $G(300\ mK) = 60$ pW/K by trimming both legs based on expected atmospheric loading and achieved optical efficiency.

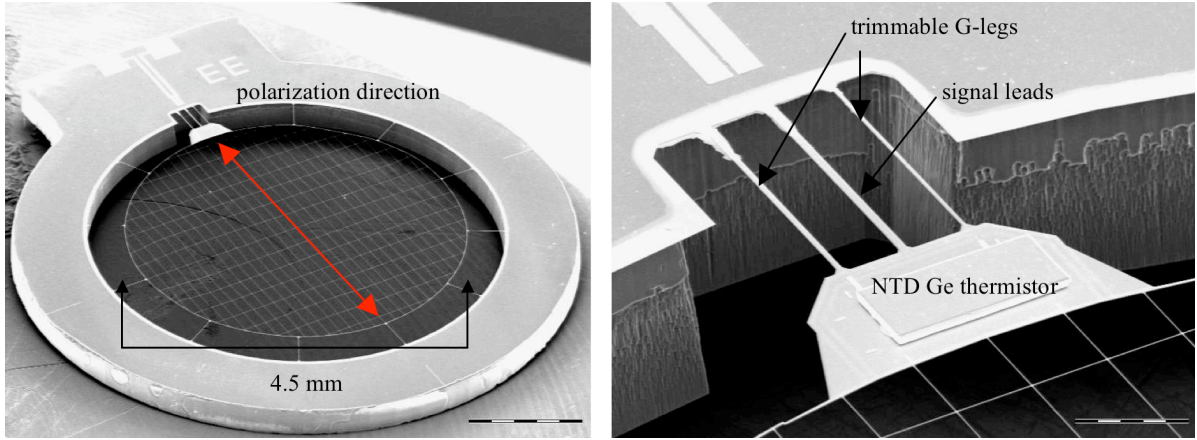

Figure 1. A Robinson $Si_3N_4$ PSB prior to being mounted and wirebonded in its module housing. Only the longitudinal legs are metalized for polarization sensitivity, and spaced at 150 $\mu$m for operation in both 100 GHz and 150 GHz bands. The metalized and trimmable pair of legs adjacent to the signal leads allows for custom tuning of the thermal conductance to the 250 mK bath, depending on expected total loading on the bolometer.

The optical time constant $\tau \sim C/G$ determines the maximum desirable scanning speed $v_{MAX} \approx \theta_{FWHM} / 3\tau_{MAX}$. Median $\tau$ was measured to be ~21 ms, with 97% of the PSBs under 45 ms. This allows for scanning at up to 5 deg/s without significant signal roll-off. Typically, the beams move across the sky at 1–2 deg/s over the elevation range of the observed field (see Section 4.2).

PSBs have been used for a number of experiments. Detectors similar in design were used in the 2003 flight of BOOMERANG at 145 GHz,[15] providing the first bolometric measurement of CMB polarization. PSB modules identical to the ones described here have been deployed in QUaD,[16] targeting CMB polarization at angular scales an order of magnitude smaller than Robinson's range. In addition, the HFI instrument of the upcoming *Planck* mission will incorporate eight PSBs (4 feeds) at each of 143, 217, and 353 GHz. The PSB concept derives from previous generations of total intensity $Si_3N_4$ micromesh bolometers,[17] used in numerous measurements of the temperature anisotropy (ACBAR, Archeops, Bolocam, BOOMERANG 98, MAXIMA and MAXIPOL).

### 2.2. Optical design

Robinson's axisymmetric refractive optical design eliminates potential polarization systematics associated with off-axis designs or on-axis secondary support structures. Because the entire optics is internal to the cryostat and cooled to cryogenic temperatures, instrument optical loading is minimized and is dominated by the ambient vacuum window. The measured upper limit on the instrument loading is < 4–5 $K_{RJ}$, consistent with the expected loading of ~3.5 $K_{RJ}$ in both frequency bands.

### 2.2.1. Zotefoam window

The 300-mm aperture vacuum window consists of four individual layers of Zotefoam PPA30,[*] heat-laminated to form a single 10-cm thick slab and sealed to an aluminum frame using Stycast epoxy. A similar low-loss Zotefoam window was used in ACBAR.[18] The window was measured to have ~1% loss at 150 GHz, adding an expected 2–3 $K_{RJ}$ of optical loading on the detectors. Scattering at mm wavelengths is very low due to the small cell size (0.3 mm) and low density (30 mg/cm$^3$). Because the foam window is radiatively cooled to below ambient temperature from both sides, frost can form from residual moisture in the air. To prevent this, we have installed a taut 18-$\mu$m polypropylene film as a cover, with the airspace between the cover and the Zotefoam window slightly pressurized and continuously purged by warm dry nitrogen gas.

---

[*] Zotefoams Inc., Walton, KY 41094. Phone: 800-362-8358. Web: http://zotefoams.com.

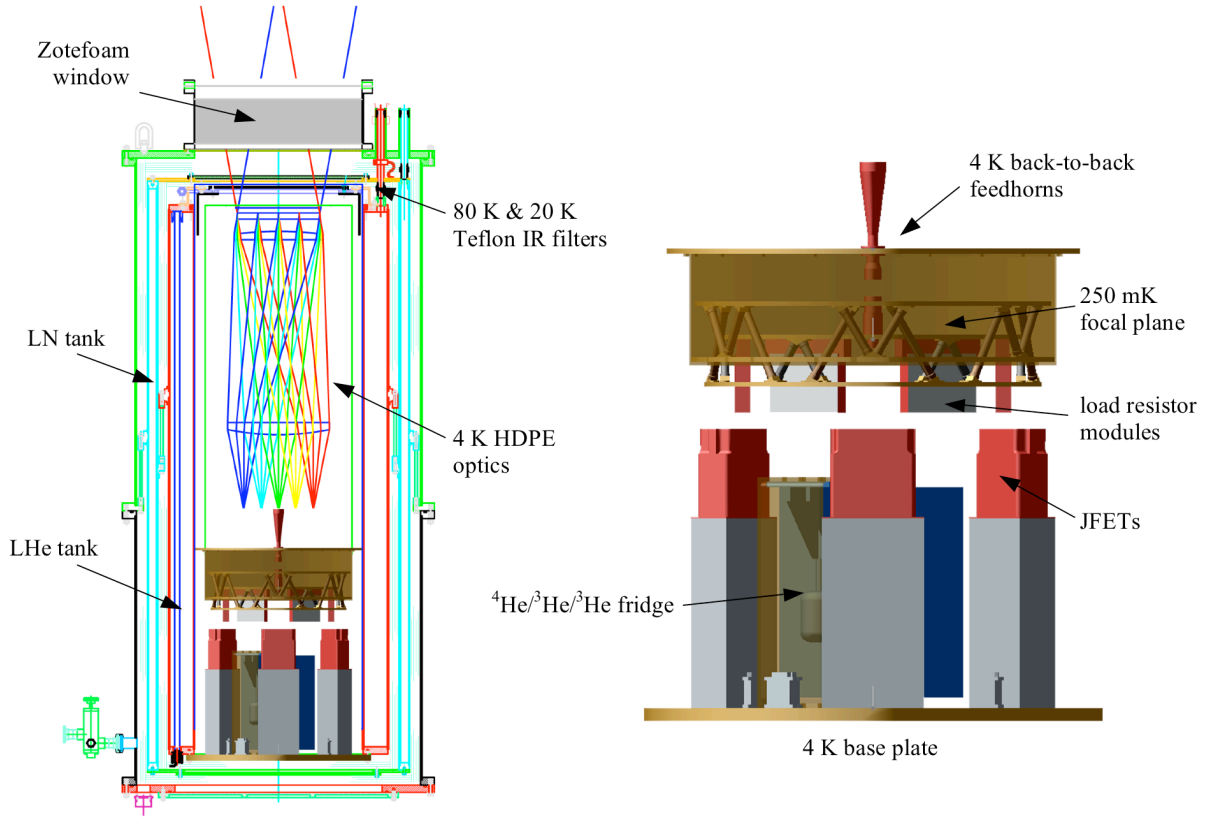

Figure 2. *Left:* A schematic diagram of the Robinson cryostat and the cold refractive optics. The LN and LHe tanks are toroidal and concentrically surround the cold optics and the receiver. Teflon IR filters at 80 K and 20 K (the latter cooled by He boil-off vapor) effectively limit the thermal loading onto the 4 K stage through the large 300-mm Zotefoam vacuum window. *Right:* The 4 K receiver provides a Faraday cage for the 250 mK PSB focal plane. For simplicity only one pixel is shown, and the cylindrical shield is omitted.

### 2.2.2. IR-blocking filters

Effectively blocking the incoming IR radiation is crucial to reducing the thermal loading onto the LHe stage and subsequent optical elements at 4 K and below. Two absorptive PTFE (Teflon) filters of thickness 1.5 cm and 2.0 cm are used at the 77 K liquid nitrogen stage and the LHe vapor-cooled 20 K stage, respectively, for this purpose.

Application as an absorptive IR-blocking filter ideally requires good in-band transmission, high thermal conductivity to dissipate absorbed radiation to the edge, ease of anti-reflection coating or low index of refraction, and effective blocking above 3 THz. Attenuation below 3 THz provides negligible mitigation of the thermal loading onto 4 K. Teflon was chosen over other candidates for meeting the requirements in all of the above areas, with low in-band loss of 0.015 Np cm$^{-1}$ at 150 GHz and a relatively low index of refraction $n = 1.44$.[19] The thermal conductivity of 1.4–2.2 mW cm$^{-1}$ K$^{-1}$ at 20–80 K is sufficient to dissipate the expected absorbed radiation,[20] and avoid excess heating of the filters beyond their usefulness. IR-attenuation properties are excellent for this application: ~3 Np cm$^{-1}$ near the cut-on region at 3 THz, and negligible transmission at higher frequencies.[21]

Left uncoated, the Teflon filters incur ~3.3% reflection per surface. Efficient anti-reflection coating is accomplished with Zitex, a porous Teflon sheet available in many thicknesses and densities. The typical relative density compared to regular Teflon is ~40%, making it ideal for coating applications in conjunction with its parent material. The measured index of refraction of $n = 1.2$ agrees with expectation.[21] A single layer of Zitex per surface is used to achieve good transmission over both 100 GHz and 150 GHz bands, with ~0.5% reflection per surface.

Due to its low melting point at 120 ºC, thin sheets of LDPE provide a cryogenically robust and mm-wave transparent bonding of the Zitex layers to the bare Teflon filters. Layers of LDPE 2-mil in thickness are sandwiched between Zitex and Teflon, and press-heated together above the melting temperature of LDPE. Teflon has a significantly higher melting point and is not degraded in this process.

With the two Teflon filters installed, the quiescent loading on the LHe is 15 L/day. This compares well with 10 L/day measured when the cryostat is blanked off.

### 2.2.3. Cold refractive optics

Robinson uses a cryogenically-cooled two-lens refractive telescope with a 17º instantaneous field-of-view. The optical design satisfies three constraints (telecentricity, imaging the focal plane to infinity, and the 250 mm entrance pupil which coincides with the top lens) with four free parameters (the axial positions and focal lengths of the two lenses). The remaining degree of freedom is chosen to minimize the optical aberrations.

For reasons of ease of manufacture and low reflection loss, we chose high-density polyethylene[*] for the lenses. The performance of the HDPE design predicted by the ZEMAX software is similar to that for silicon and fused silica designs originally considered: Strehl ratios > 0.99 and cross polarization ~few × $10^{-5}$ over the field of view. The induced linear polarization at the edge of the field is predicted to be 0.8% without the anti-reflection coatings, and less with the coatings, but this has not been confirmed with measurement.

The shape design of the lenses assumed a uniform 1.9% contraction of the lens between room and operating temperatures. The validity of this assumption was tested by measuring the $1.7 \pm 0.2\%$ contraction of a sample of the HDPE at room temperature and after immersion in liquid nitrogen. Using a Michelson interferometer, we measured an index of refraction of $n = 1.574 \pm 0.007$ at cryogenic temperatures, and we used this value in the optical design. Given an index of refraction similar to Teflon and a melting point that, at 130 ºC, tolerates the LDPE bonding technique used in AR-coating the IR-blocking filters, both sides of the bi-convex lenses were similarly coated with Zitex.

To accommodate the thermal contraction mismatch between the HDPE lenses and the aluminum enclosure, each lens is mounted on three thin vertical aluminum vanes. This design is predicted to position the lens repeatably and was observed to be stiff. Flexible braided copper straps are attached from the lenses to the 4 K base of the telescope enclosure for heat sinking, and the helium-vapor-cooled Teflon filter at 20 K minimizes the radiation incident on them. The edge of the top lens is measured to equilibrate at < 8 K so that negligible emission is expected from the lenses. Circular apertures located at the lenses and in two intermediate locations within the telescope are blackened with epoxy loaded with carbon black or iron filings[†] to minimize stray light.

### 2.3. Cryostat and $^4$He/$^3$He/$^3$He sorption refrigerator

A liquid-helium/liquid-nitrogen cryostat built by Janis Research[‡] provides the base operating temperature of 4 K. The toroidal 110-liter LHe tank conductively cools the receiver insert via a single bolt circle at its base and surrounds the refracting telescope. Parasitic losses account for 0.25 W of loading on the LHe, and an additional 45 mW from the JFETs (see Section 2.6) and 0.1 W from the large optical aperture result in a hold time of 4+ days in the field, after considering the tilt requirements of the telescope. The helium boil-off vapor is used to cool one of the two Teflon filters, intercepting infrared radiation coming through the window. In equilibrium, the perimeter of this filter is at 20 K. The long, narrow (0.25 in-diameter) copper tube which exchanges heat between the vapor and the filter holder can be susceptible to ice plugs, so special care is taken to ensure that air (nitrogen) does not flow through the tube when the cryostat is cold. The toroidal LN tank has a capacity of 55 liters and a hold time of 3 days. Cryogenic servicing for the 2006 season is scheduled at intervals of two sidereal days to coordinate with observations.

---

[*] Accurate Plastics, Inc., Yonkers, NY 10705. Phone: 800-431-2274. Web: http://www.acculam.com.
[†] Emerson & Cuming, Billerica, MA 01821. Phone: 978-436-9700. Web: http://www.emersoncuming.com.
[‡] Janis Research Company, Inc., Wilmington, MA 01887. Phone: 978-657-8750. Web: http://www.janis.com.

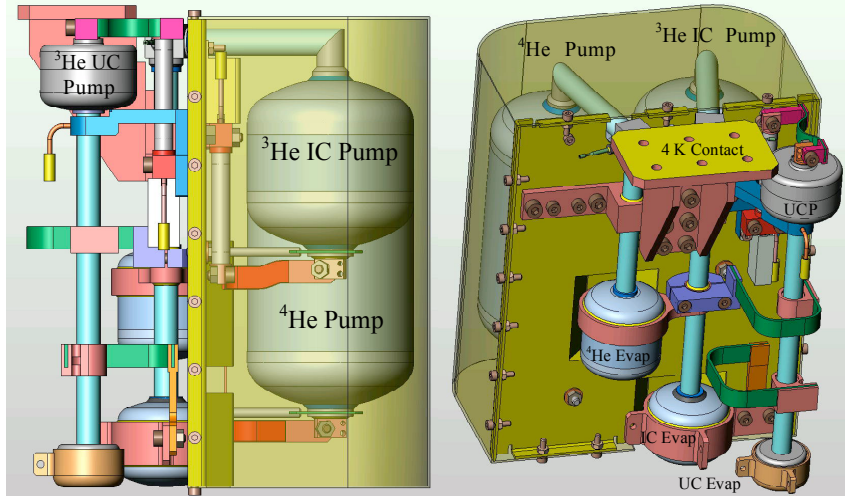

Figure 3. Diagram of the multi-stage sorption refrigerator, showing the pump and evaporator of each stage ($^4$He, $^3$He InterCooler, and $^3$He UltraCooler). The large $^4$He and $^3$He IC sorption pumps are cooled by gas-gap heat switches and housed in a light-tight shield. The small 250 mK stage UC pump is separately shielded (shield not shown). The sorption unit mounts to the main LHe tank by a copper flange (4 K contact), which connects directly to the $^4$He condensation point.

The detector focal plane is cooled to 250 mK by a 3-stage $^4$He/$^3$He/$^3$He sorption refrigerator filled with {32, 16, 2} STP liters of gas, respectively. This design eliminates the convection found in previous multi-stage designs[22] during the condensation phase[23] by putting several bends in the pump tube between the pump and the condensation point (see Figure 3). The fridge requires {4.68, 1.62, 0.41} kJ to cycle the pumps with a power dissipation of {70, 60, 60} mW to sustain the pumps at the end of the condensation phase. A similar design was used to reduce the parasitics on an adiabatic demagnetization refrigerator and a large suspended focal plane in the Z-Spec instrument.[24]

To reduce additional load on the focal plane from the un-pumped main LHe bath, we employ a thermal intercept stage in the focal plane connected to the $^4$He evaporator. This stage maintains ~1 K while the $^4$He lasts, and 2.4 K after the $^4$He is exhausted, still cooled by the $^3$He IC vapor and intercepting 32 $\mu$W of parasitic heating from the 4 K bath.

The cycling of the refrigerator is entirely automated via the control system, and requires ~5 hours to reach 250 mK from the 4 K starting point. The hold time from the start of the cycle is ~65 hours, limited by the $^3$He IC stage running out. In normal operation the refrigerator is cycled every 48 sidereal hours at the beginning of each observation schedule, providing an operating temperature duty cycle of 89%.

### 2.4. Focal plane

Each pixel on the focal plane consists of 4 K primary and re-expanding feed horns in a back-to-back configuration, followed by a small thermal gap to the 250 mK band-defining filter stack and PSB coupling feed horn (see Figure 4). This arrangement is similar to those used in ACBAR, BOOMERANG, and *Planck* HFI. The profiled corrugated primary feed horns, designed to illuminate the 250-mm aperture stop with a -20 dB edge taper in both bands, offer low sidelobe response and superior polarization characteristics over a wide bandwidth.[25] The re-expanding back-facing horns allow for quasi-free-space low pass filtering at the aperture of the PSB coupling horns. The back horns and the PSB horns are profiled in exactly the same way to aid in efficient coupling of the front-to-front apertures.

Upper edges of the two bands are defined by AR-coated multilayer low pass metal mesh filters on a dielectric substrate, heat-sunk and cooled to 250 mK. Typically, these filters exhibit spectral leaks at multiples of the cut-off frequency (3.65 cm$^{-1}$ and 5.9 cm$^{-1}$ respectively for 100 GHz and 150 GHz bands). Two additional metal mesh filters at higher cut-offs are used at each pixel to efficiently block out these leaks. Testing with thick grille filters indicates an upper limit of high frequency leak at -25 dB for 100 GHz pixels and -30 dB for 150 GHz pixels. The lower band edges are simply defined by the corrugated waveguide cut-on frequency at the throat of the PSB horn, at 83 GHz and 130 GHz.

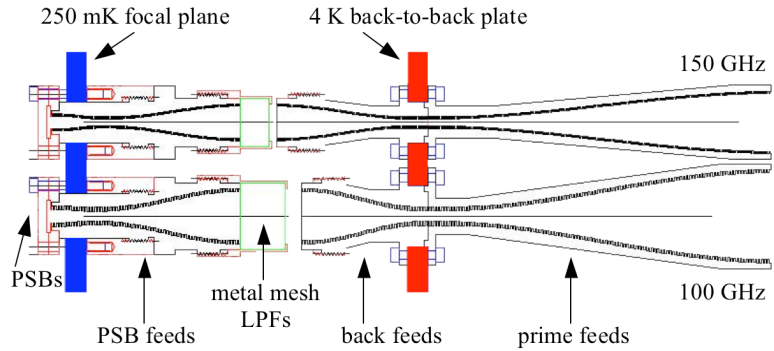

Figure 4. The optical aperture is illuminated by corrugated profiled 4 K feed horns. Re-expanding back-facing feeds provide a convenient thermal gap between the 4 K back-to-back stage and the 250 mK focal plane. Sub-K metal mesh filters and the PSB coupling feeds define the spectral bands at both frequencies.

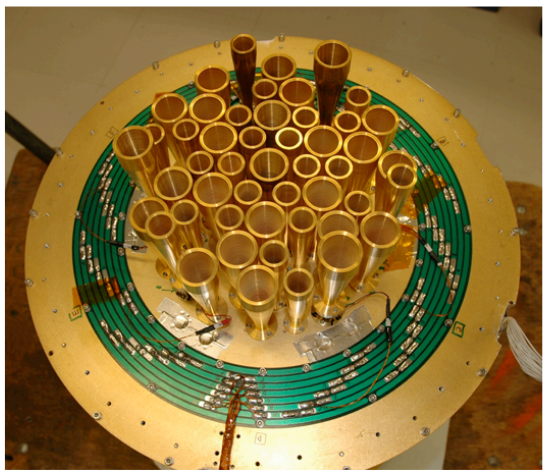
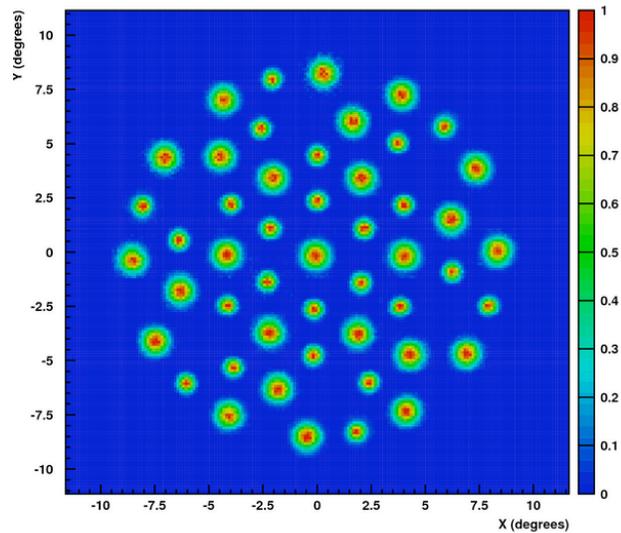

Figure 5. Fully assembled 4 K focal plane (left), and measured beams (right).

The focal plane is arranged in a six-fold symmetric pattern, as seen in Figure 5. Each hextant contains 4 pixels at each frequency. The central 100 GHz pixel is read out by one of the six hextants, accounting for a total of 96 + 2 light bolometers. Additional readout channels provide for one pair of 5 MΩ resistors, dark bolometers, and high sensitivity NTD thermistors in each hextant for diagnosing systematics.

Each PSB pair can be installed in either boresight *"Q"* or *"U"* orientation, defined with respect to the radius from the center of the focal plane. For the inaugural observing season, we have chosen to alternate between the two orientations in adjacent hextants, such that upon odd-multiple 60º rotation of the instrument about the boresight, we achieve complete parity in *Q/U* coverage on the sky.

A PCB fan-out board on the backside of the detector focal plane routes the individual PSBs of each hextant to the six load resistor modules (LRM) on the perimeter for readout by the JFET amplifier stage. The entire focal plane, from the 4 K back-to-back throat section down to the output of the JFET modules, resides within a tightly-sealed Faraday cage, eliminating any stray radiation coupling to the bolometers.

## 2.5. Faraday rotation modulators

Six of the 49 pixels deployed for this first season of observation are employed as testbeds for Faraday rotation modulators[12,26] (FRM), operated as polarization switches in a phase-sensitive detection scheme between the primary and back feed horns. Modulation offers the ability to minimize susceptibility to systematic instrumental polarization, tune the post-detection audio band for more flexibility in avoiding microphonic lines, and mitigate optical sources of systematic polarization such as differential gain fluctuations. Modulation increases the system's immunity to offset variations downstream of the primary feed horns.

The FRMs are functionally similar to a rotating birefringent half-waveplate, but have no moving parts and are implemented at the back-to-back feed horn section of each pixel. Modulation is achieved using the Faraday effect, whereby the plane of linearly polarized radiation is rotated during propagation through a magnetized dielectric. A superconducting solenoid, biased with ±0.1 A, produces ±45º rotation, allowing measurement of both $Q$ and $U$ Stokes parameters with a pair of orthogonal PSBs. Rotation angle has been measured to be uniform across both of the optical bands and is extremely stable over time. Current FRM performance meets desired design criteria: ~80% transmission, ~1% reflection, ~1% instrumental polarization, < 1% cross polarization, > 90% polarization efficiency, and < 1 mW rms total power dissipation. More information on the FRM design and performance is in preparation.[27]

## 2.6. Readout electronics

The readout electronics system uses digitally-generated AC bias current, cold JFETs, analog preamplifiers, and digital demodulators. The behavior of the bias generator and digital electronics can be changed on demand from the control software, enabling, for instance, DC mode operation for measuring load curves.

A sinusoidal 100 Hz AC bias current, applied symmetrically across 2 × 20 MΩ load resistors to minimize common mode pickup, eliminates 1/f electronics noise above 10 mHz. The quasi-stationary optical signals can then be recovered using synchronous demodulation of the AC signals. The bolometer bias is produced by a highly stable, digitally-controlled sinusoidal waveform synthesizer capable of varying the bias frequency from 100 to 200 Hz.

The bolometer readout front end consists of cold JFET amplifiers operated at ~120K, mounted to the 4 K baseplate and connected to the 250 mK LRM via low thermal conductivity manganin twisted-pair cables.[*] The JFETs are necessary to amplify the signal from high impedance bolometers and protect the bolometer from EMI. Three JFET modules (developed originally for the Herschel/SPIRE instrument) with two sets of membranes each service the six hextants, dissipating a total of 45 mW at 4 K. The best 24 channels out of 30 available are selected from each membrane, achieving an average voltage noise contribution of 7 nV/√Hz. The output of the JFET modules exits the 4 K Faraday cage through RF-filtered connectors.

Upon exiting the cryostat, the signal is routed inside an RF-tight cage to the room-temperature preamplifiers, each equipped with a commandable high pass filter for AC bias mode operation. The preamplifier output is then anti-alias filtered and exits the "clean" electronics through RF-filtered connectors to be digitized synchronously with the bolometer bias. In AC bias mode, the digitized signal is multiplied by synchronous sine and cosine, averaged and filtered by a gaussian digital finite impulse response filter in a low power mixed signal processor. In DC bias mode, the digitized signal is instead multiplied by ±1. This allows for the demodulator transfer function to be measured by applying a delta function in DC test mode, if needed.

The demodulated sine and cosine signals from all the bolometers are packaged onto an ethernet bus, converted into an optical fiber signal and sent to the acquisition computer. The optical fiber data are routed through the central section of a pair of slip rings[†] to allow for unimpeded boresight rotation of the cryostat.

## 2.7. Telescope mount

The telescope mount addresses the requirements of reliable winter-over operation at the South Pole by enclosing all serviceable components of the system except for the receiver window itself within an easily accessible shirt-sleeve environment. Its unique design, illustrated in Figure 6, accomplishes this while allowing steering and tracking of the telescope through a full range of motion in azimuth (400º), elevation (50–90º), and continuous rotation about the boresight, or theta axis.

---

[*] Tekdata Interconnect Systems, Staffordshire, ST6 4HY, UK. Phone: +44 (0) 1782 254700.
Web: http://www.tekdata-interconnect.com.
[†] Moog Components Group, Blacksburg, VA 24060. Phone: 800-336-2112. Web: http://www.polysci.com.

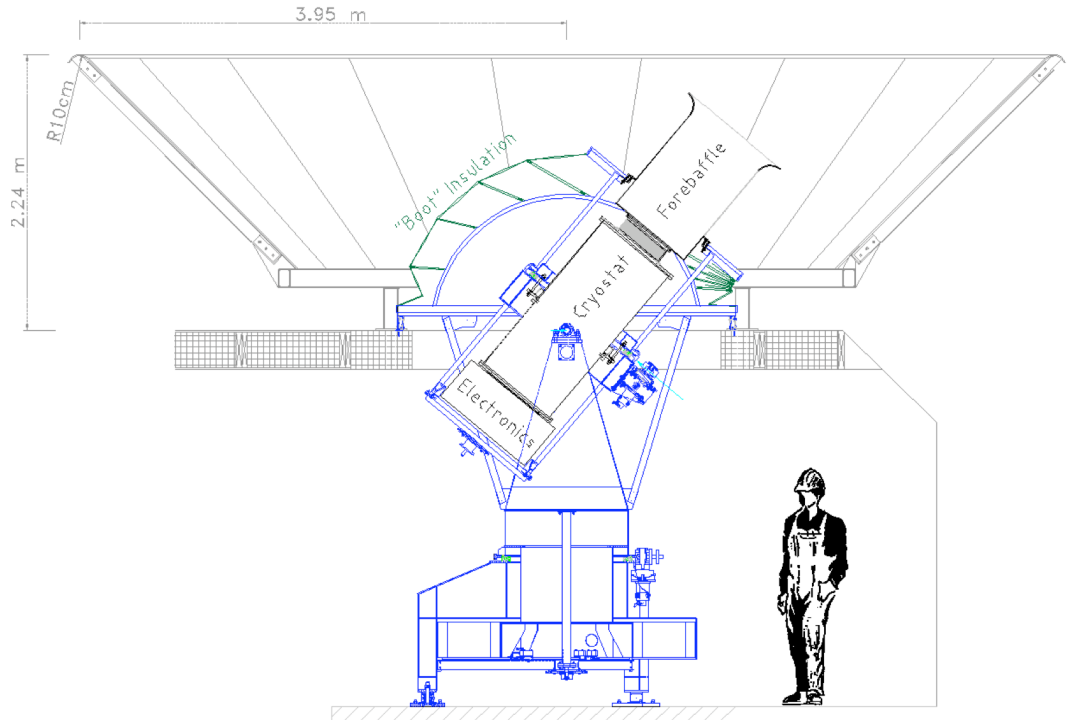

Figure 6. The three-axis telescope mount, cryostat, and ground shield as installed on site. The entire cryostat and the receiver electronics reside inside the DSL building, with only the optical aperture and the forebaffle exposed to the elements. A flexible and co-rotating "boot" insulation allows for unhindered movement of the telescope.

The mount is situated in the upper floor of the new Dark Sector Laboratory (DSL), 0.8 km from the geographic South Pole. An insulated fabric bellows permits motion in the elevation axis, and together with double brush seals at the azimuth and theta axes forms an environmental seal between the rooftop of the building and the front of the receiver which maintains the space around the cryostat, control electronics, and drive assemblies at room temperature. The enclosed observatory space is kept at a slight positive pressure, so that a constant outward airflow through the brush seals and special vents directed toward the window eliminates ice buildup there.

The Robinson mount was engineered and fabricated in conjunction with TripointGlobal/VertexRSI.[*] Lightweight box steel construction lends the mount extreme rigidity and immunity to flexure; the combined weight of the telescope, when fully equipped and operational, is approximately 7,500 lbs. It is supported on a steel and wood platform attached to the structural beams of the building. Continuous tilt monitoring and periodic star pointing have confirmed short term stability and blind pointing accuracy of the combined mount/platform structure to meet our pointing spec of < 20″. Long term drifts at the level of ~1′ per month appear to be dominated by settling of the building.

The mount allows rapid scanning, up to 5 deg/s, about the azimuth and/or theta axes while maintaining precise pointing and producing minimal vibration. These axes employ ultra-quiet crossed roller bearings[†] and gearless cycloidal motor reducers.[‡] Integrated testing with the Robinson receiver drove a choice of toothed belt drives[§] for the azimuth and theta axes, which further reduced high drive speed microphonic excitation to levels well below our detector noise floor.

---

[*] VertexRSI, Kilgore, TX 75662. Phone: 903-984-0555. Web: http://www.tripointglobal.com.
[†] ROLLON Corp., Sparta, NJ 07871. Phone: 877-976-5566. Web: http://www.rollon.com.
[‡] Gates Mectrol Corp., Salem, NH 03079. Phone: 603-890-1515. Web: http://www.mectrol.com.
[§] TB Wood's Inc., Chambersburg, PA 17201. Phone: 717-264-7161. Web: http://www.tbwoods.com.

## 2.8. Optical star pointing camera

We aim to achieve overall pointing accuracy better than ~1% of the beam size to limit contamination to the B-mode polarization. This requires a precise knowledge of the dynamic state of the mount, including flexure, axis tilts, and encoder offsets. To aid in the pointing reconstruction, we have built an optical star-pointing refractor camera with a 2″ resolution, mounted adjacent to the main optical window and co-aligned with the boresight rotation axis.

Since there are ten dynamic parameters (including the collimation error of the pointing telescope itself), a complete calibration requires observation of at least 20 stars. To be able to establish a pointing model during the Antarctic summer, the camera was designed to be sensitive enough to detect magnitude 3 stars in daylight. For maximum contrast against the blue sky, a sensor with enhanced near-IR sensitivity is used with an IR72 filter.[*] The 100-mm diameter lens, color-corrected and anti-reflection coated for 720–950 nm, was designed by Anthony Stark of the Smithsonian Astrophysical Observatory for the South Pole Telescope (SPT). Its 901-mm focal length results in a small 0.5° field of view. Careful adjustments of the CCD camera and the mirrors reduced the star camera's collimation error to 2.4´. During the austral summer season prior to the current observing campaign, we were able to successfully capture stars down to magnitude 2.8 for initial pointing calibration.

## 2.9. Ground screen and forebaffle

We use two levels of shielding against ground radiation contamination: an absorptive forebaffle fixed to the cryostat and a large reflective screen fixed to the roof of DSL. The geometry of the shielding is such that any ground radiation must be diffracted at least twice before entering the window in any telescope orientation during operation.

The forebaffle is an aluminum cylinder lined with a microwave absorber to minimize reflected radiation into the telescope. It is sized to clear the sidelobes of the edge pixels, and long enough so that the window will never directly see either the outer ground screen or that of the proposed SPT adjacent to the Robinson facility. The dimension of the forebaffle also prevents moonlight up to 21° elevation from entering the window directly. The Moon is above this elevation 5 days a month. The forebaffle's aperture lip is rounded with a 13-cm radius to reduce diffraction.

After testing many materials for the forebaffle absorber, we chose a 10-mm thick open-cell polyurethane foam sheet,[†] which had the lowest measured reflectivity (< 3%) at 100 and 150 GHz[28] when placed on a metal surface. To prevent snow from accumulating on the porous Eccosorb foam, it is lined with 1.6-mm thick smooth Volara polyethylene foam. The combined Eccosorb HR/Volara stack was measured to reflect ~5% at 100 GHz. In addition, a self-regulating heat cable is wrapped around the outside of the baffle to sublimate any snow on the baffle surface, if necessary. The additional loading on the bolometers due to this forebaffle was measured to be ~1 $K_{RJ}$. Since the baffle is fixed with respect to the detectors, its thermal emission is not expected to significantly affect the differential measurements. The attenuation of the ground pickup by the forebaffle was measured on site to be > 10–20 dB below the already low sidelobes of the primary beams, limiting the far-sidelobe response to < -30 dBi beyond 30° off-axis.

The 2-m tall outer screen reflects any stray sidelobes to the relatively homogenous cold sky. The 8-m top diameter is sized so that the diffracted ground radiation will never directly hit the window even at the low end of our observing elevation at 55°. As with the forebaffle, the edge of the outer screen is rounded with a 10-cm radius to reduce diffraction.

## 3. SOUTH POLE SITE

The National Science Foundation Amundsen-Scott Station is located at the geographic South Pole, on the interior ice plateau of Antarctica at an altitude of 2800 m. As was detailed in Keating *et al.* 2003 and elsewhere,[29] the sky above the South Pole is extremely dry and stable, resulting in 150 GHz opacity of $0.03 < \tau < 0.04$ and atmospheric noise contamination far below Robinson's instrumental noise at least 80% of the time. We use a 350-$\mu$m tipper to continuously monitor the atmospheric transmission, independent of once-a-day sky dip measurements performed at the beginning of every observation cycle. Extrapolating the 350-$\mu$m tipper measurements for the month of May 2006 to 150 GHz, we observe {25, 50, 75}% opacity quartiles of {0.034, 0.036, 0.037}.

---

[*] AVA Astro Corp., Hudson Falls, NY 12839. Phone: 877-348-8433. Web: http://www.astrovid.com. (Astrovid StellaCam EX CCD)
[†] Emerson & Cuming, Randolph, MA 02368. Phone: 781-961-9600. Web: http://www.eccosorb.com. (Eccosorb HR)

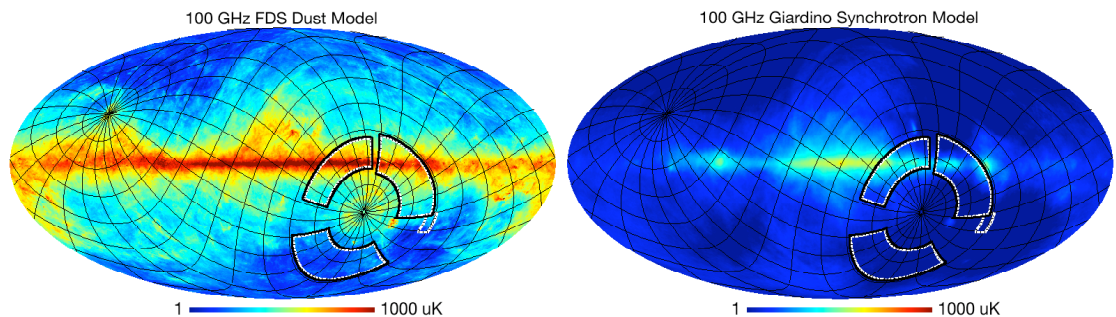

Figure 7. Robinson spends 18 hours/day observing the low galactic foreground CMB region (lower annulus outlined in black). An additional 6 hours every other day are devoted to the Eta Carina galactic region (upper right annulus). A brighter galactic plane region (upper left annulus) is used for short calibration scans. The coverage region of QUaD, our sister experiment, is outlined in white for reference.

In addition to the excellent atmospheric conditions, from the South Pole it is possible to track a given field continuously at constant elevation angle. The sky that is accessible for observation from the Pole includes some of the largest regions of minimal galactic foreground emission on the celestial sphere. These characteristics of the Pole make it the ideal location for the long, deep integrations needed to study the CMB polarization signal. Robinson is designed to exploit these advantages in the course of extremely long integrations on selected fields. The design of the telescope also reflects the particular environmental challenges of South Pole observations. During the six months of the year for which the Sun (the strongest source of contamination) is below the horizon, presenting optimal observing conditions, the average ambient temperature is -60°C. Very long integration times demand a design that can ensure reliable telescope performance during this period.

## 4. OBSERVATION STRATEGY

We have chosen a ~1000 deg$^2$ region with low expected dust and synchrotron emission for our primary CMB science field. In addition, we devote a small but significant amount of observing time on a portion of the galactic plane using the exact same scan strategy as the CMB field to study the foreground polarization in detail. Our operational cycle is based around a 48 sidereal-hour schedule, during which we cycle the sub-K fridge and service the cryogens, cover the galactic region once, and map the CMB region twice.

### 4.1. Observed fields

The mount is capable of observing down to a boresight elevation of 50º, providing access to the minimal foreground regions accessible from the South Pole. As illustrated in Figure 7, we have selected the region centered at (RA = 0 hr, dec = -57.5º) as the main science field, where we spend a total 18 hours/day. A secondary field centered at (RA = 9 hr, dec = -57.5º), encompassing the bright Eta Carina galactic plane region, is observed for 6 hours every other day, complementing our primary science target with a deep study of the galactic polarization. In addition, a third region at (RA = 15:42 hr, dec = -55.0º), in a significantly brighter part of the galactic plane, serves as a daily calibration source, and we perform repeated fixed-elevation scans for 10 minutes at the beginning and end of every 9-hr observing block.

### 4.2. Scan strategy

Fixed-elevation scans at 2.8 azimuth deg/s provide the primary signal modulation. This scan rate maps power on the sky on the angular scales of interest ($l \sim 30$–$300$) to time domain signals in the range of 0.1–1.0 Hz, above any residual $1/f$ fluctuations. For a given field, we perform a raster scan, stepping up in elevation by 15´ at the end of every 50 minutes of azimuth scan and eventually covering 5 degrees in boresight elevation, from 55º to 60º. We have chosen not to track the field center in RA while scanning; rather, we scan about a fixed azimuth and update the tracking to catch up with the field once per elevation step. This has the advantage of making AZ-synchronous and scan-synchronous signals to be degenerate with each other, and separable from the signal on the celestial sphere. We can, if we so choose, project out a model of the AZ/scan-synchronous signal from each scan in RA in the final analysis. This approach is superior to a simple lead-trail field differencing in preserving information on angular scales corresponding to $l > \sim 40$. There is no penalty associated with the fixed-AZ strategy if we decide against the baseline subtraction.

Each 48-hr observing block is broken up into four distinct phases. The first 6 hours are spent cycling the sub-K fridge back to 250 mK and servicing the cryogens. Then an 18-hr block of CMB observation completes the first day of the block. The beginning 6 hours of the second day are used to observe the galactic plane region, then another 18-hr CMB scan follows. The order in which the upper and lower halves of the elevation range are covered for the CMB field is swapped between the first and second days, allowing a jackknife to be performed to check for any time-variable AZ-fixed contamination. Each 48-hr block is executed at one of four instrument boresight orientations, at $\theta = \{-45º, 0º, 135º, 180º\}$, giving two independent and complete $Q/U$ coverages of the field. With 78% scan turn-around efficiency and actual CMB observing of 16.7 hours/day, the total observing efficiency is ~54% for the CMB science region.

### 4.3. Gain calibration

Multiple levels of absolute and relative gain calibrations are integrated into the observing strategy. In addition to the galactic calibration scans that bookend every 9-hr block of observation, each 50-minute constant-elevation scans is bookended by elevation nods and IR flash calibrator pulses. Absolute gain and polarization calibration using a dielectric sheet is performed on a regular basis throughout the winter observing season.

Absolute temperature scale can also be reliably obtained by cross-calibrating between Robinson and WMAP in either map or Fourier space. Preliminary results based on two months of data are shown in Section 5.

#### 4.3.1. Dielectric sheet calibrator

To characterize the polarization response of the PSBs, we use a dielectric sheet calibrator based on POLAR's design,[30] providing an absolute temperature calibration to ±10% and polarization orientation to within ±1° on a monthly basis.

A small partially polarized signal of known magnitude is created by using an 18-$\mu$m polypropylene film as a 45° beam splitter in front of the telescope aperture, reflecting a fraction of the beam onto an ambient load. The ambient load lining the inside surface of the cylindrical calibrator is exactly analogous to the Eccosorb HR/Volara layers used in the forebaffle, giving ~95% emissivity over the two bands and presenting a similar optical load as during normal observing. This particular setup produces a partial polarization of amplitude ~100 mK at 100 GHz and ~250 mK at 150 GHz, which is small enough to ensure that the change in bolometer responsivity due to the calibration process itself is negligible (see Figure 8).

During calibration, the forebaffle is replaced with the dielectric sheet calibrator. Boresight rotation of the cryostat while pointed at zenith modulates the partially polarized signal induced by the dielectric sheet. Unlike POLAR's single on-axis beam, Robinson's off-axis beams see complicated but calculable deviations from the simple sinusoidal modulation.

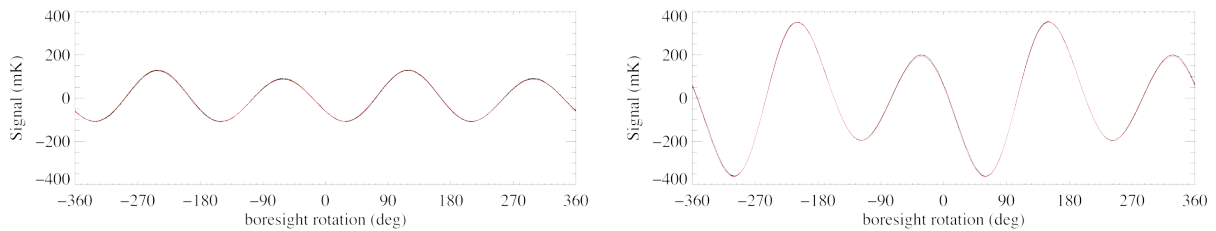

Figure 8. Dielectric sheet calibrator response (black) and model fits (red) for two PSB pair. *Left:* 100 GHz *"Q"* pixel, 4.1° off-axis from boresight. *Right:* 150 GHz *"U"* pixel, 8.5° off-axis. Orthogonal PSBs within a pixel are differenced to reject common-mode drift during the measurement. Absolute temperature calibration is determined to ±10% and polarization orientation to within ±1°.

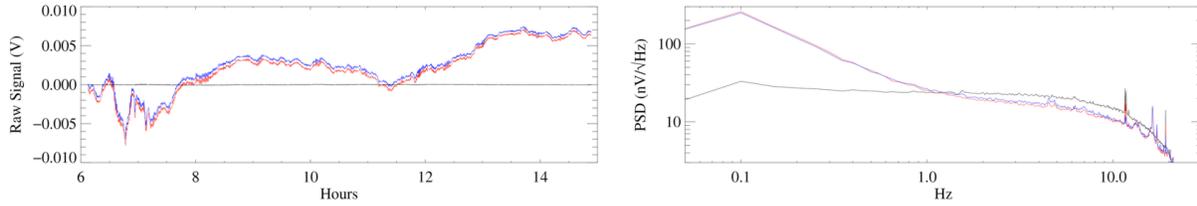

Figure 9. Relative gain-adjusted PSB pair differencing is able to remove common-mode atmospheric fluctuations down to 0.1 Hz, corresponding to $l \sim 20\text{–}35$. Shown in red and blue are individual 150 GHz PSBs within a pair, and in black is the pair differenced. A single relative gain fit over a period of 9 hours was made based on the fluctuations alone, but in practice, the relative gain is measured twice every 50 minutes using both elevation nods and IR flash calibrator source.

We have performed two calibrations so far this season, and both times we were able to fit the measured signal to within a few percent of the theoretical model and derive the PSB orientations to within 1°. The uncertainty in the absolute gain is expected to be ~8%, mostly due to ~3% uncertainty in the polypropylene film thickness.

### 4.3.2. Elevation nods

At the beginning and end of every 50-minute fixed-elevation scans across the science target, we perform "elevation nods" for the purpose of tracking the relative response of the bolometers. An elevation nod consists of a sinusoid-like "up-down-return" (normal) or "down-up-return" (inverse) elevation motion of the telescope with total amplitude of 1 degree and duration of 1 minute. We measure $dV_i/dA$, the voltage response of each bolometer $i$ to the change in airmass, to correct for the different radiation response of the bolometers. The elevation nods are centered on the same elevation as the contemporaneous science azimuth scan, and they produce a signal of 10–20 mK across the focal plane for a typical atmospheric zenith loading of 10 K. The atmosphere is believed to have negligible linear polarization.[31,32]

The primary purpose of the elevation nods is to determine the relative response of the bolometers within a PSB pair so that atmospheric emission and CMB temperature anisotropy are rejected by the bolometer difference (see Figure 9). Over a month of observation, we have measured a typical repeatability of 0.6% (1-$\sigma$) in the gain ratio $(dV_{i,front}/dA)/(dV_{i,back}/dA)$, after subtracting systematic trends with atmospheric loading. The elevation nods serve an additional purpose of measuring the relative gains of the detectors over the focal plane, thereby allowing the construction of co-added maps from all of the pixels at a given frequency.

In principle, atmospheric emission is rejected perfectly by the elevation nod technique since an atmospheric signal is used to measure the gains. However, since the CMB has a different spectral slope within our bands compared to the atmosphere, good rejection of CMB temperature anisotropy requires good spectral matching of the bolometers. From model atmospheric spectra and typical measured spectral mismatch, we estimate that temperature anisotropy is rejected by a factor of $10^2$ for a single PSB. Rejection is improved by an additional factor of ~10 since the map pixels are visited by multiple detectors with multiple rotation angles during the scan.

The elevation nod motion produces a thermal disturbance of the focal plane that is detected as a false $dV/dA$ in dark bolometers at a level 0.3% of the response in the light bolometers. Alternating between normal and inverse elevation nods allows us to partially subtract this effect. Without accounting for the thermally-induced response and spectral matching, we expect residual CMB temperature anisotropy to appear in our polarization maps at the ~0.1 $\mu$K level.

### 4.3.3. IR flash calibrator

In addition to the elevation nods, we have implemented an infrared source as an independent relative calibrator. An electrically modulated 2.25 mm² infrared emitter packaged into a compact collimating optics[*] is embedded at the end of a Zotefoam arm on the side of the BICEP window. The Zotefoam arm swings into the center of the beam on demand from the control system. Once in the beam, the IR source is pulsed at 1 Hz for a period of ~1 min, producing a typical optical signal of ~100 mK. The flash calibration precedes every elevation nod for redundancy. We have found that the reliability with which the relative gain is determined compares favorably with that of the elevation nod method, giving us added leverage and redundancy in dealing with potential systematic issues in gain measurements.

---

[*] Hawkeye Technologies LLC, Milford, CT 06460. Phone: 203-878-6892. Web: http:// http://www.hawkeyetechnologies.com. (Part no. IR-55)

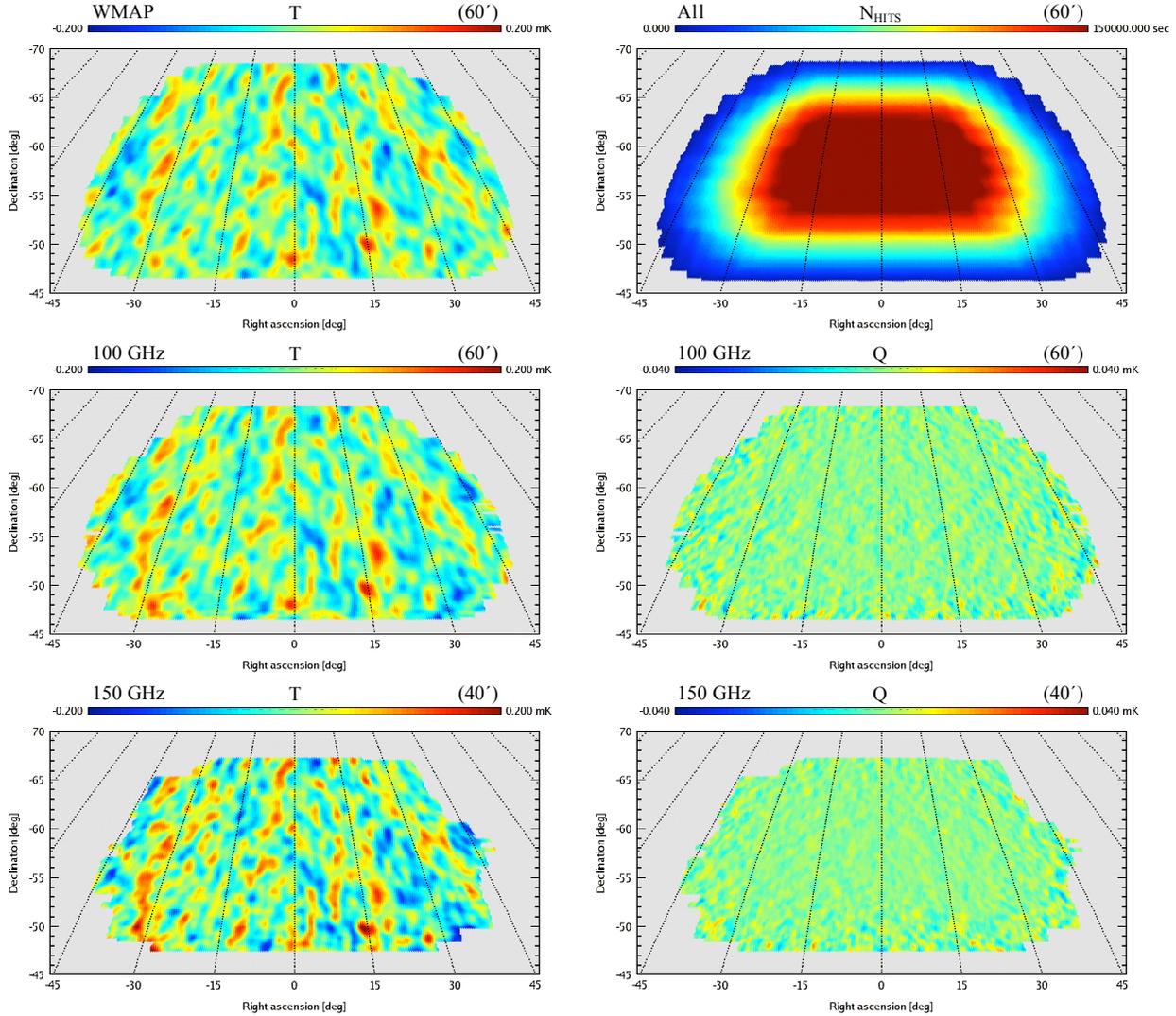

Figure 10. Preliminary maps from the first two months of observations (March and April 2006). *Top Left:* WMAP $T$ map, W-band (94 GHz), filtered with the Robinson 100 GHz observing strategy. *Middle Left:* Robinson $T$ map, 100 GHz channels. The good agreement with WMAP confirms the ability of our detector pairs to measure the common mode $T$ signal over atmospheric noise, and provides a cross-calibration. *Bottom Left:* Robinson $T$ map, 150 GHz channels. *Top Right:* Map of total integration time per 1 deg$^2$ pixel at 100 GHz and 150 GHz, smoothed with the beam profile. *Middle Right:* Robinson Stokes $Q$ map at 100 GHz. The color scale for the $Q$ maps shows a narrower range by a factor of 5 in comparison to the $T$ maps. In the deepest central region (600 deg$^2$), the rms in 1 deg$^2$ pixels is 2.0 $\mu K$, consistent with the expected noise level. *Bottom Right:* Robinson Stokes $Q$ map at 150 GHz. In the deepest central region (400 deg$^2$), the rms in 1 deg$^2$ pixels is 1.5 $\mu K$, consistent with the expected noise level.

# 5. PRELIMINARY RESULTS

We have acquired and analyzed two months of preliminary data taken during March and April 2006, prior to fine-tuning of our scan strategy and proper RF shielding of the warm electronics. The analysis was made with blind pointing, and relative gains were derived from elevation nod bolometer response versus expected airmass for a given pixel position. Raw data are processed with rudimentary deglitching and Fourier band pass filtered (0.1–5.0 Hz for $T$ and 0.2–5.0 Hz for $Q/U$) prior to noise-weighted naive map-making.

The lower four panels of Figure 10 show beam-smoothed $T/Q$ maps for 100 GHz and 150 GHz channels from the two months of data. We image the degree angular-scale CMB temperature anisotropy with high signal to noise on a month-timescale. March versus April jackknife shows that the $Q$ and $U$ (not shown but statistically similar to $Q$) maps are consistent with noise at the ~2.0 $\mu K_{CMB}$ (100 GHz) and ~1.5 $\mu K_{CMB}$ (150 GHz) per 1 deg$^2$ pixel level, in line with expectation given the integration time.

We expect the approximate absolute calibration of the temperature responsivity derived from the el-nods to be accurate to ~20%. Using the high S/N temperature map, however, we can attempt a cross-calibration of the 100 GHz map with WMAP for a more accurate calibration of the temperature scale. Using Robinson's pointing information over March/April, a timestream was simulated from WMAP's 3-yr W-band temperature map. The simulated timestream was identically filtered and naively binned as Robinson's data. The derived map from WMAP (top left panel of Figure 10) and the Robinson 100 GHz map are in excellent agreement. With the WMAP cross-calibration in place, we confirm the measured median NET$_{CMB}$ for 100 GHz and 150 GHz to be 480 $\mu K \cdot s^{1/2}$ and 420 $\mu K \cdot s^{1/2}$, respectively.

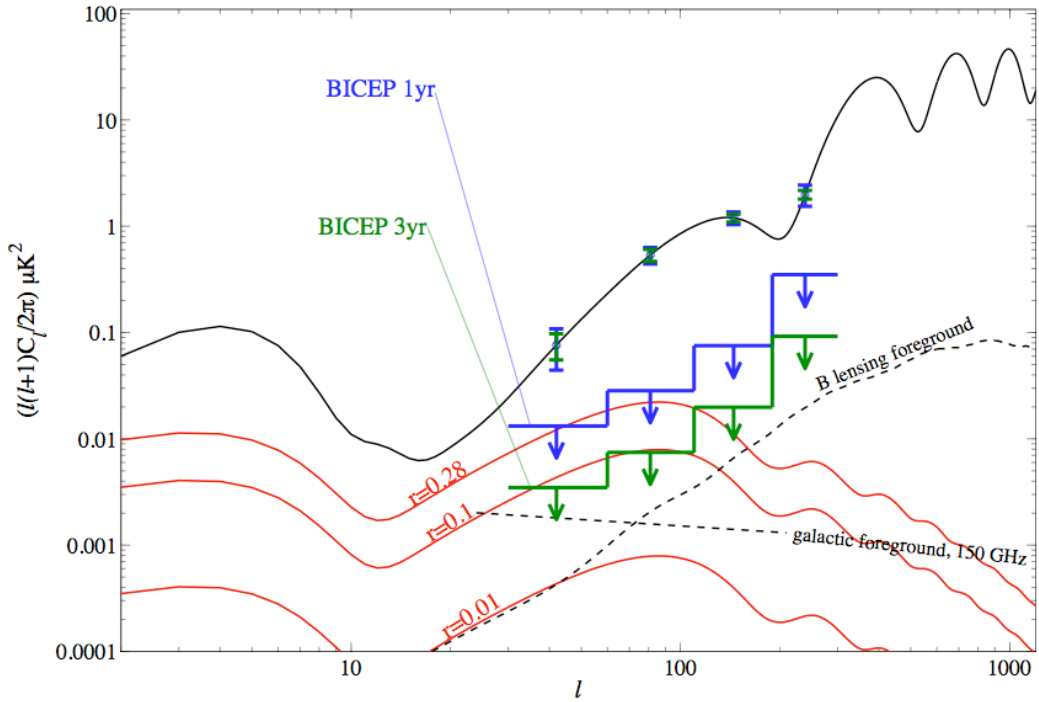

Figure 11. CMB polarization spectra and B-mode sensitivity levels (1-$\sigma$ in band power) for 1 season (100 days) and 3 seasons (380 days) of observing. The top curve shows CMB E-mode polarization, generated by density perturbations. The three lower curves show B-mode polarization from inflationary gravitational waves at the level of the best current upper limits ($r = 0.28$) and in the range $r = 0.1$–0.01 that is suggested by the simplest models of inflation. The dominant B-mode foregrounds, which arise from gravitational lensing of E-mode polarization and from galactic emission present in the observed field, are shown as dashed lines at the levels of best current estimates.

## 6. EXPECTED SENSITIVITY

Given the WMAP-calibrated NETs and measured observing efficiency, we estimate the expected sensitivity. Figure 11 shows the projected sensitivities to both E-mode and B-mode polarization, assuming 100 days of good observing in the present first season, and a total of 380 good days at the end of thee seasons. Effects of E/B separation are included in these calculations, but not that of foreground removal.

Because the Robinson CMB field was chosen for extraordinarily low dust emission, polarized dust and synchrotron are estimated to have roughly equal levels at 150 GHz, though these levels have at least a factor of 2 uncertainty in the upward direction and still greater uncertainty in the downward direction. Estimated levels of dust and synchrotron polarized emissions were calculated from extrapolating the FDS[33] (model 8) and WMAP K-band[9] $Q/U$ datasets, respectively, to 150 GHz for our given field of observation. The dust extrapolation assumes ~5% polarization. We expect to challenge the best upper limits on the B-modes by the end of this season, and push well past it in the coming years of additional observations and approach levels of greatest cosmological interest.

## 7. CONCLUSION

The Robinson Telescope, a degree angular scale bolometric polarimeter on the BICEP experiment, was successfully deployed for its first season of observation at the South Pole in early 2006. The instrument, including its cryogenics, receiver, and control subsystems, has performed exceedingly well, and we have already collected a wealth of high-quality data in the early months of the winter observing season. Based on the initial analysis of early data, we have made a number of precautionary changes in our scanning and operational strategies to further minimize potential scan-synchronous contaminations. Gain calibration and PSB pair differencing, critical to faithful measurement of the CMB polarization, are found to be functioning well at the current level of noise reached. We expect continued, uninterrupted operation through October 2006. The instrument is currently funded to observe through the 2007 season, and a proposal is pending to extend observations with Robinson through 2008, and to significantly upgrade the BICEP experiment with a larger format array of detectors for 2009.

## ACKNOWLEDGEMENTS

Robinson has been made possible by generous support from NSF Grant No. OPP-0230438, Caltech President's Discovery Fund, Caltech President's Fund PF-471, and JPL Research and Technology Fund. BGK also gratefully acknowledges support from NSF CAREER Award #0548262. In addition, we are especially indebted to the late J. Robinson whose thoughtful gift was instrumental in the continued progress of this project. The success of this instrument from conception to deployment rested heavily on the capable staff of machinists and engineering staff at Caltech, UC Berkeley, Stanford, and UC San Diego; the advice and helpful discussions from our colleagues in ACBAR, BOOMERANG, QUaD, and Bolocam; and logistical and administrative support from Kathy Deniston. We also thank the Planck/Herschel SPIRE project for use of their JFET testbed, Todd Gaier for his assistance in VNA testing of the feed horns, and Stanford for providing the load resistor modules.